\documentclass[journal]{IEEEtran}
\usepackage{graphicx}
\begin{document}
\title{Single Photon Source Driver Designed in ASIC}
%
%

\author{Bo Feng, Futian Liang, Xinzhe Wang, Chenxi Zhu, Yulong Zhu, and Ge Jin
\thanks{Manuscript received June 2, 2018. This work was supported by the National Natural Science Foundation of China under Grants No. 61401422.}
\thanks{Bo Feng, Xinzhe Wang, Yulong Zhu, and Ge Jin are with State Key Laboratory of Particle Detection and Electronics, University of Science and Technology of China, Hefei, Anhui 230026, P.R. of China (e-mail: fengbo@mail.ustc.edu.cn, wxzyf@mail.ustc.edu.cn, yl9194@mail.ustc.edu.cn, goldjin@ustc.edu.cn).}%
\thanks{Futian Liang is with Hefei National Laboratory for Physical Sciences at the Microscale and Department of Modern Physics, University of Science and Technology of China, Hefei, Anhui 230026, P.R. of China,
and Chinese Academy of Sciences (CAS) Center for Excellence and Synergetic Innovation Center in Quantum Information and Quantum Physics, University of Science and Technology of China, Shanghai 201315, P.R. of China (email: ftliang@ustc.edu.cn).}%
\thanks{Chenxi Zhu is with School of Microelectronics, University of Science and Technology of China, Hefei, Anhui 230026, P.R. of China (e-mail: chenxi61@mail.ustc.edu.cn)}%
\thanks{First author: Bo Feng, Corresponding author: Futian Liang.}%
}

\maketitle
\thispagestyle{empty}

\begin{abstract}
The single photon source is an important part of the quantum key distribution (QKD) system.
At present, the single photon source is large in size and complex in structure for a lot of discrete components which are used.
The miniaturization of the photon source is the tendency of the QKD system.
We integrate all laser driver electronic module into one single ASIC chip, which can be used to drive the 1550nm DFB laser in random pulse mode and it can greatly reduce the volume of the single photon source.
We present the design of the chip named LSD2018 and simulation results before the tape-out. The LSD2018 is fabricated with a 130 nm CMOS process and consists of a discriminator, an adjustable pulse generator, a bandgap reference, an SPI bus, and an amplitude-adjustable current pulse driver. The electronic random pulse from the driver can go 20mA to 120mA in amplitude and 400ps to 4ns in pulse width. The parameters can be set by an SPI bus.
\end{abstract}


\section{Introduction}
\IEEEPARstart{t}{he} QKD system has been proved to be unconditionally secure by the uncertainty principle and the no-cloning theorem in quantum mechanics.
\cite{gisin2002quantum}\cite{takesue2007quantum}
The ideal choice in the QKD system is the true single photon source.
However, the suitable deterministic single photon source is still not available. \cite{rusca2018finite}
We use the phase-random weak-coherent light emitted by the distributed feedback (DFB) laser as a source.
To meet the needs of the QKD system for the source, the DFB laser should be precisely modulated.
Therefore, a precise laser driver circuit is necessary.
The 1550nm DFB laser needs a drive current pulse signal with the frequency up to 625MHz, the pulse width from 400ps to 800ps, and the amplitude from 20mA to 100mA.
Because of the fast speed and the high current of the drive signal, most of the single photon source drivers we use consist of a lot of discrete components.
For the better performance, the design of the circuit board must be very compact.
To make the driver more integrated, we designed an ASIC chip named LSD2018, a laser source driver chip used to drive the 1550nm DFB laser in random pulse mode.

\begin{figure}[!t]
\centering
\includegraphics[width=3.5in]{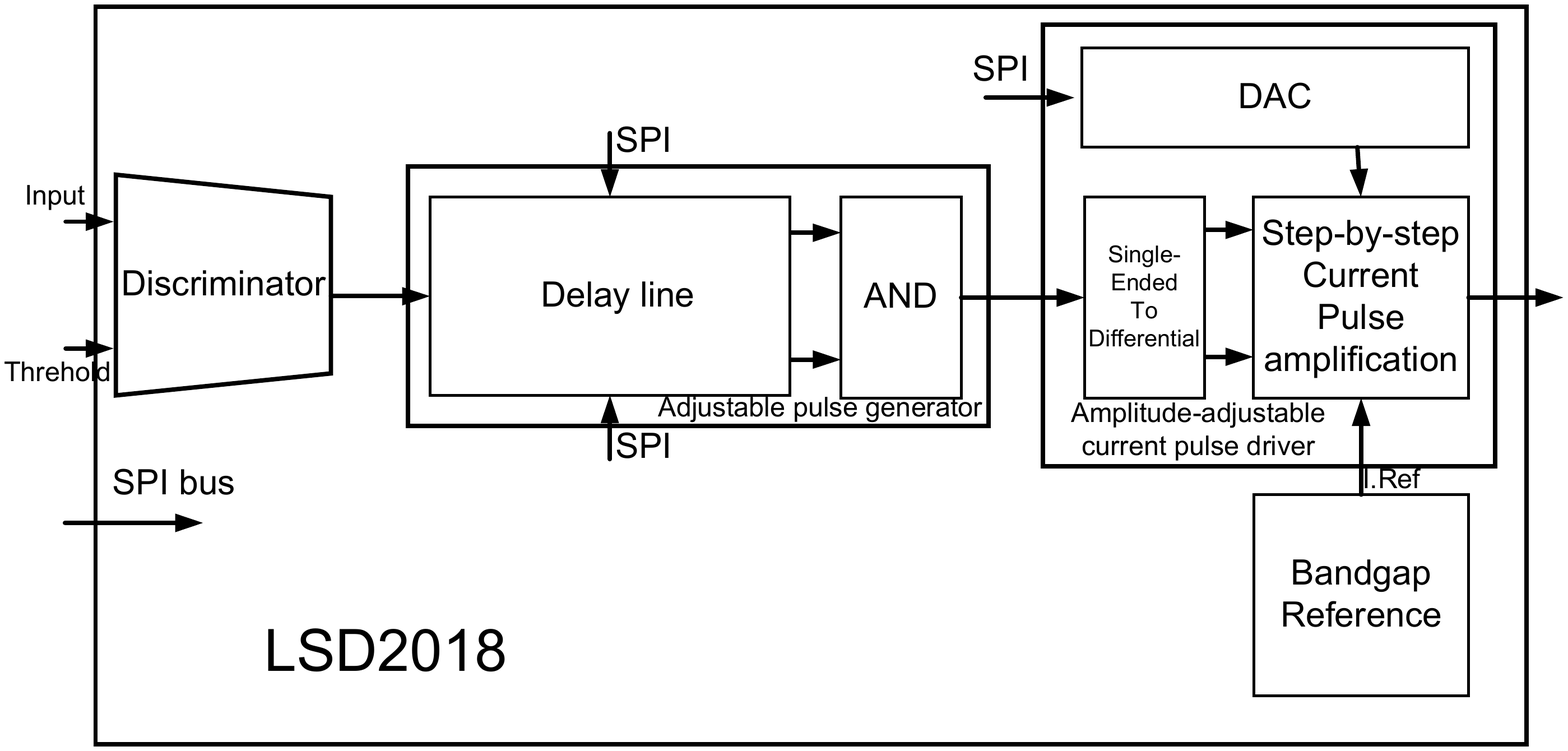}
\caption{The structure of the LSD2018}
\label{fig_structure}
\end{figure}

\section{Design Scheme}


The structure of the LSD2018 is shown on Fig. \ref{fig_structure}.
The LSD2018 consists of a discriminator, an adjustable pulse generator, a bandgap reference, an SPI bus, and an amplitude-adjustable current pulse driver.
The discriminator compares the input signal and the threshold and outputs a signal with the amplitude from 0 to 1.2V.
The bandgap reference provides the reference current for the entire circuit.
The SPI bus delivers the control signals which are used to configure the adjustable pulse generator and the amplitude-adjustable current pulse driver.

\subsection{Adjustable Pulse Generator}


The adjustable pulse generator consists of a delay line and an AND gate.
The generator receives the signals from the discriminator and generates two time-delay signals by the delay line.
The two time-delay signals are different in delay time and phase.
So, when the two signals are added by the AND gate, we can get a narrow pulse signal.
The pulse width of the narrow pulse signal is determined by the relative delay of the two time-delay signals.
The delay of the two time-delay signals is configured by the SPI bus.

\subsection{Amplitude-adjustable Current Pulse Driver}


The amplitude-adjustable current pulse driver consists of a single-ended to differential module, a step-by-step current pulse amplification module, and a 4-bit DAC.
The single-ended to differential module receives the output signal from the adjustable pulse generator and converts it from a single-ended signal to a differential signal.
The step-by-step current pulse amplification module converts the differential signal into a current pulse signal and amplifies the current amplitude step by step.
Finally, it generates a signal with high current amplitude up to 120mA to drive the laser. \cite{tao2003low}
The amplitude of the output current can be configured by the DAC.
The DAC is configured by the SPI bus.

\begin{figure}[!t]
\centering
\includegraphics[width=3.5in]{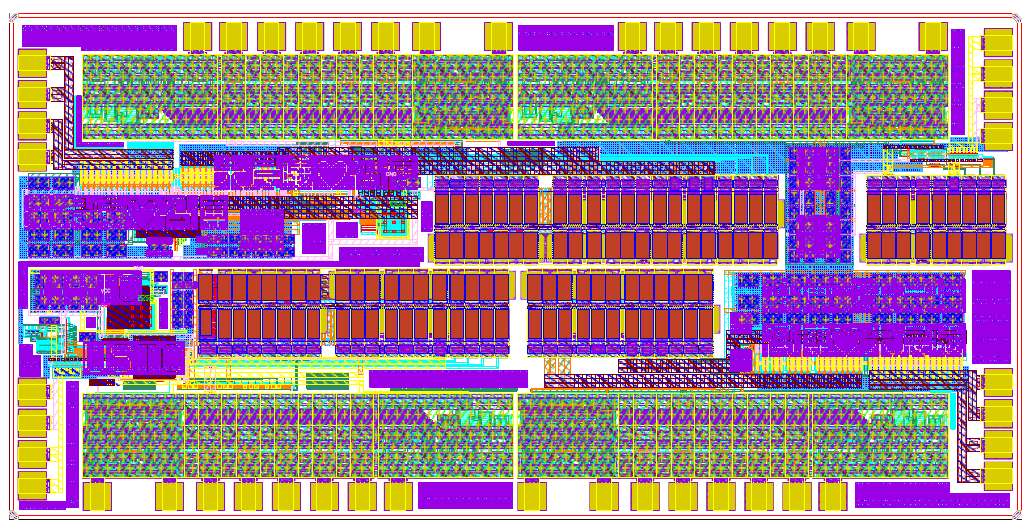}
\caption{The layout of the LSD2018}
\label{fig_layout}
\end{figure}

The LSD2018 is designed in a 130nm CMOS process.
The die size is 2.4mm $\times$ 1.2mm and consists of 4 different drivers.
They are different from each other in parameters and integration.
So that, we can test more easily and comprehensively.
The layout of the LSD2018 is shown in Fig. \ref{fig_layout}.


\begin{figure}[!t]
\centering
\includegraphics[width=3.5in]{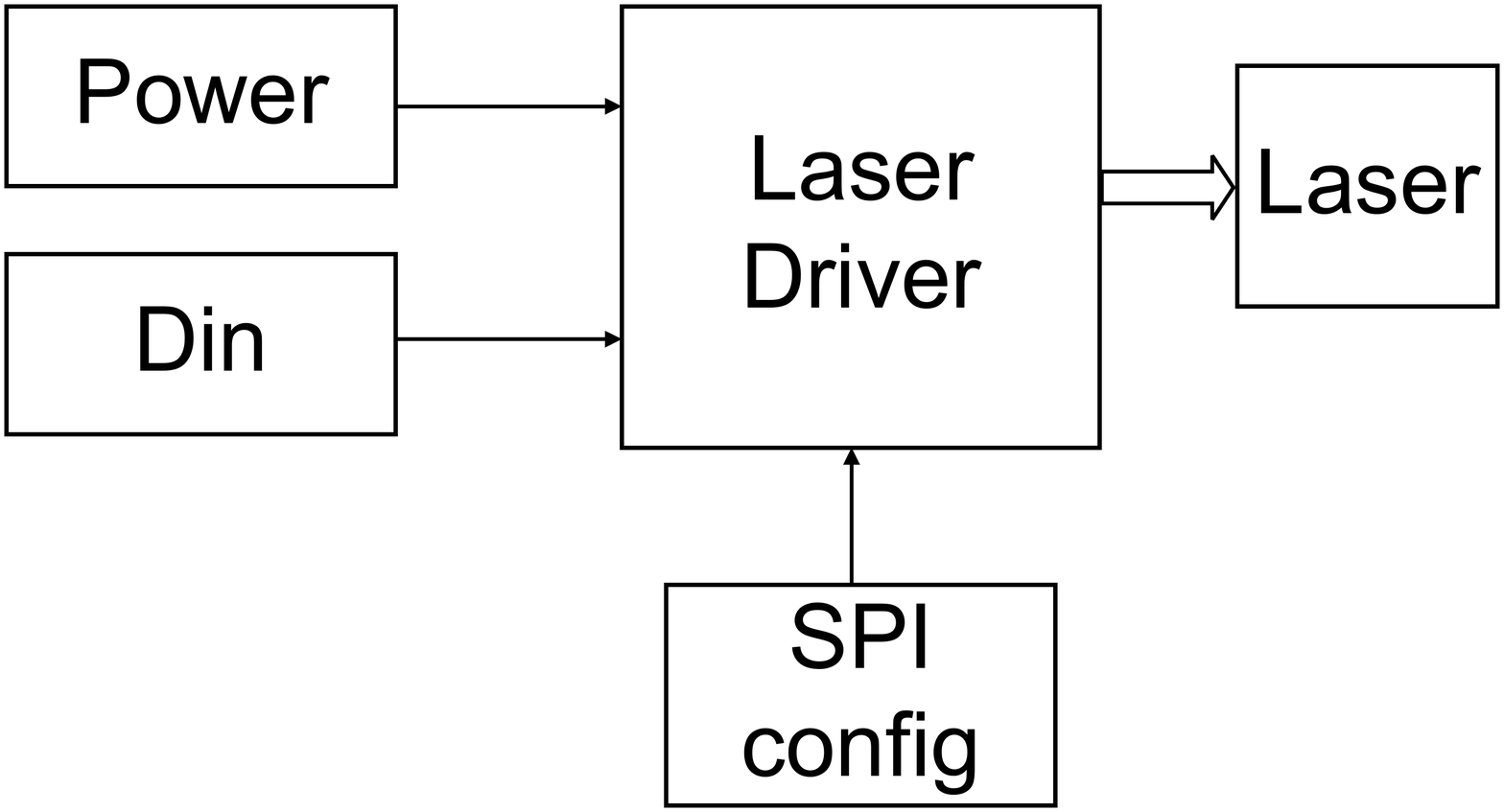}
\caption{The layout of the simulation}
\label{fig_sim}
\end{figure}

\section{Simulation}



The LSD2018 tape-outed on May 15, 2018.
The schedule delivery date is August 2018.
In the paper, we provide simulation results only.
We simulated across different process corners (tt, ss, ff) and at different temperatures (0$^\circ$C, 27$^\circ$C, 85$^\circ$C).
In the simulation on layout in different conditions, the LSD2018 is fully functional.
The structure of simulation is shown in Fig. \ref{fig_sim}.
Considering the influences of the parasitic parameters, some resistances, capacitances and inductances are connected in the rail of power and the line of input and output.
We place a pull-up resistor at the output as the 1550nm DFB laser.

\begin{figure}[!t]
\centering
\includegraphics[width=3.5in]{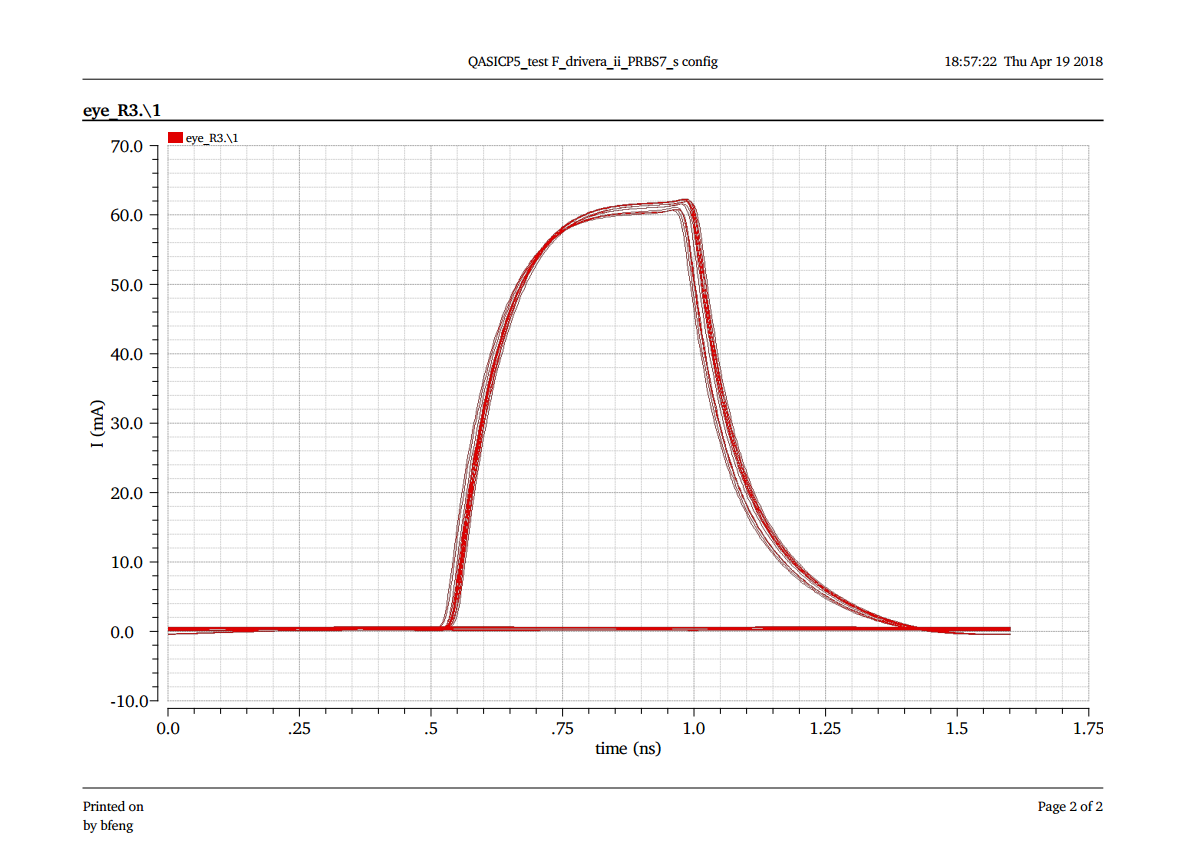}
\caption{ the ``eye diagram" of the simulation which the output current frequency is 625MHz, the pulse width is 400ps, and the amplitude is 60mA at the process corner tt and 27$^\circ$C
(Because the output current signal is a pulse signal which only consists of long 0 and short 1, the ``eye diagram" looks incomplete.)}
\label{fig_eye}
\end{figure}

The Fig. \ref{fig_eye} is the ``eye diagram" of the simulation which the output current frequency is 625MHz, the pulse width is 400ps, and the amplitude is 60mA at the process corner tt and 27$^\circ$C.
The off current is less than 0.6mA.
The time jitter is about 20ps on the rising edge, and about 30ps on the falling edge.
The range of the amplitude change is about 2mA at 60mA.
The design indicators and the simulation results are shown in Table \ref{table_LSD}.
The LSD2018 can generate a current pulse signal of which the frequency is up to 625MHz, the amplitude is from 20mA to 120mA and the pulse width is from 400ps to 4ns.
It can satisfy the requirements of the 1550nm DFB laser.

\begin{table}[!t]
\renewcommand{\arraystretch}{1.3}
\caption{The design indicators and the simulation results of the LSD2018}
\label{table_LSD}
\centering
\begin{tabular}{|c|c|c|}
\hline
  & Project indicators & simulation results\\
\hline
Maximum output frequency & 625MHz & 625MHz\\
\hline
Minimum output pulse width & 400ps & 400ps\\
\hline
Output current amplitude & 20mA-100mA & 20mA-120mA\\
\hline
\end{tabular}
\end{table}

\section{Conclusion}


In the primary simulation, the LSD2018 is fully functional.
It can generate a drive current pulse signal with the high current from 20mA to 120mA, the narrow pulse width from 400ps to 4ns, and the fast speed up to 625MHz.
The drive signal can ideally drive the 1550nm DFB laser.
By integrating laser driver electronic module into one ASIC chip, the LSD2018 greatly reduces the volume of the single photon source driver.
It is the beginning of the miniaturization of the QKD system.
The full performance tests will be done when the LSD2018 is delivered.

\end{document}